\documentclass[11pt,a4paper]{article}
\pdfoutput=1
\usepackage{microtype}
\DisableLigatures[f]{encoding = *, family = * }
\usepackage{upref,amsmath,amssymb,amsthm,mathrsfs,epsf,amsfonts}
\usepackage{graphicx}
\usepackage{hyperref}
\usepackage{paralist}
\usepackage{multirow}
\usepackage{float}
\usepackage{url}
\usepackage{rotating}
\usepackage{epstopdf}
\usepackage{tablefootnote}
\usepackage{xfrac}
\usepackage{paralist}
\usepackage[round]{natbib}
\usepackage{caption}
\usepackage{subcaption}
\usepackage[table]{xcolor}
\usepackage[version=0.96]{pgf}
\usepackage{tikz}
\usetikzlibrary{arrows,shapes,automata,backgrounds,petri}
\usepackage{authblk}
\usepackage[section]{placeins}

\usepackage{setspace}
\renewcommand{\arraystretch}{1.1}

\newcommand*{\equal}{=}

\newcommand{\mv}[1]{{\boldsymbol{#1}}}

\newcommand*{\abs}[1]{\left\lvert#1\right\rvert}

\newcommand*{\md}{\,\mathrm{d}}
\newcommand*{\bmat}[1]{%
\begin{bmatrix}#1\end{bmatrix}}

\newcommand*{\proper}{\mathsf}
\newcommand*{\pP}{\proper{P}}
\newcommand*{\pE}{\proper{E}}

\newcommand{\Normal}[2]{\proper{N}\left(#1,#2\right)}
\newcommand{\GammaDist}[2]{\proper{\Gamma}\left(#1,#2\right)}
\newcommand{\Dir}[2]{\proper{Dir}\left(#1,#2\right)}
\newcommand{\BetaDist}[2]{\proper{Beta}\left(#1,#2\right)}
\newcommand{\IW}[2]{\proper{IW}\left(#1,#2\right)}
\newcommand{\Cauchy}[2]{\proper{C}^+\left(#1,#2\right)}

\graphicspath{{./Figures/}}

\begin{document}
\title{Reconstruction of Past Human land-use from Pollen Data and Anthropogenic land-cover Changes Scenarios}

\author[1,2,3]{Behnaz Pirzamanbein}
\affil[1]{\small Department of Applied Mathematics and Computer Science, Technical University of Denmark, Denmark}
\affil[2]{\small Centre for Mathematical Sciences, Lund University, Sweden}
\affil[3]{Centre for Environmental and Climate Research, Lund University, Sweden}
\author[2]{Johan Lindstr\"{o}m}
\renewcommand\Authands{ and }
\date{\vspace{-5ex}}
\maketitle

\begin{abstract}
  Accurate maps of past land cover and human land-use are necessary when
  studying the impact of anthropogenic land-cover changes on climate. Ideally
  the maps of past land cover would be separated into naturally occurring
  vegetation and human induced changes, allowing us to quantify the effect of
  human land-use on past climate. Here we investigate the possibility of
  combining regional, fossil pollen based, land-cover reconstructions with,
  population based, estimates of past human land-use. By merging these two
  datasets and interpolating the pollen based land-cover reconstructions we aim
  at obtaining maps that provide both past natural land-cover and the
  anthropogenic land-cover changes.

  We develop a Bayesian hierarchical model to handle the complex data, using a
  latent Gaussian Markov random fields (GMRF) for the interpolation. Estimation of
  the model is based on a block updated Markov chain Monte Carlo (MCMC)
  algorithm. The sparse precision matrix of the GMRF together with an adaptive
  Metropolis adjusted Langevin step allows for fast inference. Uncertainties in
  the land-use predictions are computed from the MCMC posterior samples.
  
  The model uses the pollen based observations to reconstruct three composition 
  of land cover; Coniferous forest, Broadleaved forest and Unforested/Open land. 
  The unforested land is then further decomposed into natural and human induced 
  openness by inclusion of the estimates of past human land-use. The model is
  applied to five time periods - centred around 1900 CE, 1725 CE, 1425 CE, 1000
  and, 4000 BCE over Europe. The results suggest pollen based observations can be
   used to recover past human land-use by adjusting the population based 
   anthropogenic land-cover changes estimates.
\end{abstract}

\newpage


\section{Introduction}
\label{secD:intro}

Human activities mainly influences the climate through the emission of
greenhouse gases and anthropogenic land-cover changes (ALCC)
\citep{KalnaC2003_423}. The effects of both natural and human induced land-cover
changes on climate have been investigated in several simulation studies at both
global 
\citep[e.g.][]{claussen2001biogeophysical, brovkin2002carbon,
  bala2007combinedclimate, betts2007biogeophysical, pitman2009uncertainties,
  pongratz2009effects, christidis2013role, Armstrong2016roleCO2}
and regional scales \citep[e.g.][]{KalnaC2003_423, strandberg2013regional}.

Historic ALCC consists mainly of deforestation to allow for
agriculture and urbanization \citep{ruddiman2005humanimpact}. The temperate
latitudes simulation studies indicate that replacing forests with agricultural
land tends to decrease the radiative forcing (and thus temperature)
\citep{bala2007combinedclimate, betts2007biogeophysical}, while observational
studies show local temperature increases due to urbanization
\citep{KalnaC2003_423}. The temperature decreases due to human deforestation
are, to some extent, balanced by greenhouse gas 
emission due to the deforestation ($\text{CO}_2$) and farming practices
(Methane) on the deforested land \citep{ruddiman2005humanimpact,
  kaplan2013forest}. Earth system models that include dynamic vegetation,
allowing for feedback between changes in climate, global $\text{CO}_2$-levels,
and vegetation, give an even more complex
picture. For these models the effects of ALCC depends on the global
$\text{CO}_2$-levels, the climate region, and the natural land-cover replaced by
human land-use \citep{Armstrong2016roleCO2}.

Comparing historical temperature records with past natural land-cover and ALCC
might improve our understanding of interactions among climate, 
land cover, and human land-use \citep{strandberg2013regional}. However,
descriptions of both past natural land-cover
\citep[e.g.][]{brovkin2002carbon, strandberg2011high, hickler2012projecting}
and past ALCC scenarios
\citep[e.g.][]{kaplan2009prehistoric, pongratz2009effects, klein2011hyde}
varies considerably \citep{gaillard2010holocene}. It was previously shown
that fossil pollen records can be used to reconstruct past vegetation and land
cover at both local \citep{sugita2007Love}, regional \citep{sugita2007Reveals,
  paciorek2009mapping, sugita2010testing}, and continental scales
\citep{Pirzamanbein2014}.

This paper investigates the possibility of reconstructing both past
natural land-cover and the ALCC by extending the Bayesian hierarchical model
introduced by \citet{pirzamanbein2018modelling}. The fossil pollen data can be used to
obtain past land cover \citep{sugita2007Reveals}, but does not distinguish
naturally open land from deforestation caused by ALCC. Ideally we would like to
combine land-cover estimates based on fossil pollen records with
archaeological data. However, initial studies of available archaeological data
revealed a number of potential issues (see discussion in
Section~\ref{secD:archaeological}).

To investigate if the modelling is possible we instead used
ALCC scenarios \citep{kaplan2009prehistoric, klein2011hyde} as an
estimate of past ALCC. The resulting model can be seen as an adjustment of the
ALCC scenarios based on information in the pollen records (The available data is
described in Section~\ref{secD:data}). The reconstruction is done across Europe
for five time periods --- 
centred around 1900, 1725, 1425 CE and 1000, 4000 BCE. These time periods
represent important historical periods (recent past, little ice age, black death,
late bronze age, and early Neolithic) and are commonly used in both climate
modelling and palaeoecological studies. In Section~\ref{secD:archaeological} we
outline one way of extending the model to include archaeological data, and
we hope that our results will encourage the development of archaeological
databases, that can be used in future modelling.

The model presented here (see\ Section~\ref{secD:model}) considers the pollen
based land-cover data to be Dirichlet observations and the ALCC scenarios to
beta observations of underlying latent fields. The spatial structure in the
latent fields is modelled using covariates and Gaussian Markov Random Fields
\citep{lindgren2011explicit}. The model is estimated using a Markov chain Monte
Carlo (MCMC) algorithm based on the Metropolis Adjusted Langevin algorithm
(MALA) \citep{MALA}. 
Results are presented in Section~\ref{secD:results} and
Section~\ref{secD:conclusion} concludes the analysis with a discussion.

\section{Data} \label{secD:data}
The available data consist of fossil pollen based land-cover data, estimates of
past human land-use (ALCC scenarios) and potential covariates (elevation and
output from a dynamic vegetation model -- DVM).

\subsection{Pollen based land-cover compositions}
\label{subsecD:Rev}
Pollen based estimates of three land-cover compositions (LCCs), Coniferous
forest (C), Broadleaved forest (B) and Unforested land (U), were obtained
from the LANDCLIM project \citep{gaillard2010holocene} using the
REVEALS model \citep{sugita2007Reveals}. These three land-cover types are commonly
used in studies of past climate and climate modelling
\citep{strandberg2013regional}. REVEALS is a mechanistic model which uses
inter-taxonomic differences in pollen productivity, dispersal and the size of
sedimentary basins to estimate regional land-cover from pollen 
records. The sedimentary pollen records used by REVEALS are obtained from
lakes and bogs and presented as grid based REVEALS estimates for the $1^\circ\times
1^\circ$ grid cells containing sampled lakes and/or bogs \citep[][showed that the spatial scale of REVEALS reconstructions is around $100\times100$ km]{hellman2008effects}. The resulting land-cover data consists of pollen
based LCCs for respectively 175, 181, 193, 204 and 196 grid cells during the
five time periods centred around 1900, 1725 and 1425 CE, 1000 and 4000 BCE
\citep{Trondman2015}. For use in climate modelling these sparse LCC observations
can be interpolated to continuous spatial maps \citep{pirzamanbein2018modelling}. Here we
will perform the interpolation while also trying to separate the LCC into natural
vegetation and ALCC.

\subsection{Anthropogenic land-cover change scenarios}
Two anthropogenic land-cover change (ALCC) scenarios are used as estimates of
human land-use: \begin{inparaenum}[1)] \item The Kaplan and Krumhardt 2010 scenario \citep[KK10;][]{kaplan2009prehistoric}, and \item The History Database of the Global
  Environment \citep[HYDE; ][]{klein2011hyde}\end{inparaenum}. KK10 and HYDE
are both based on historic human population density estimates, the land needed
to feed that population, and soil productivity. To match the pollen records,
the two estimates of human land-use were upscaled (by averaging) to the
$1^\circ\times 1^\circ$ grid cells.

\begin{figure}[htp]
  \includegraphics[width=\textwidth]{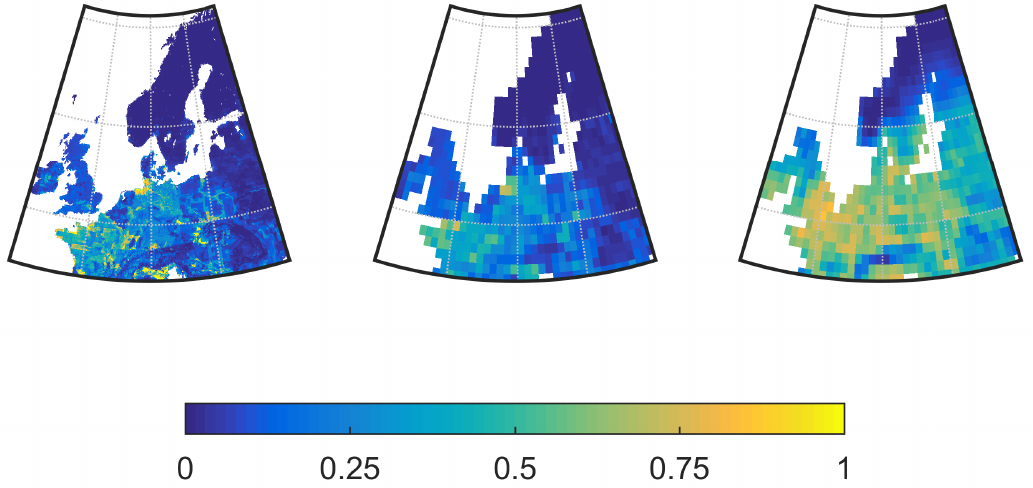}
  \caption{Anthropogenic land-cover changes (ALCC) scenarios for 1400 CE. From left to
    right: The high-resolution ($5^\prime$ or about $10$ km) HYDE ALCC scenario
    \protect{\citep{klein2011hyde}}, HYDE upscaled to $1^\circ$ resolution
    matching the pollen data, and the KK10
    \protect{\citep{kaplan2009prehistoric}} ALCC scenario at $1^\circ$ resolution.}
  \label{figD:hyde_1400}
\end{figure}

The KK10 and HYDE datasets differ substantially for the older time periods (see
Fig.~\ref{figD:hyde_1400}), due to differences in assumptions, modelling
approaches, and historical records used. In general KK10 gives
higher estimates of human land-use. Both datasets exhibit substantial local
structure.

\subsection{Covariates}
To capture large scale structures in the LCC, covariates consisting of
elevation \citep[from the Shuttle Radar Topography Mission\footnote{downloaded
    from \url{ftp://topex.ucsd.edu/pub/srtm30_plus/} on 2011--09--03},][]{elevation} 
and model based vegetation estimates can be used \citep{pirzamanbein2017bayesian}.

The model based estimates of potential natural vegetation were obtained by
running a process-based dynamic vegetation model (DVM), LPJ-GUESS,
\citep{smith2001representation} for the study area and specified time
periods. LPJ-GUESS estimates the potential natural vegetation based on
bio-climatic variables such as temperature, precipitation, and soil types
\citep[see][for details regarding the LPJ-GUESS runs]{Pirzamanbein2014}. 

\section{Model} \label{secD:model}
For the modelling we assume that each grid cell has a natural LCC, $\mv{p}_L=
(p_C,p_B,p_U)$, representing the proportion of each grid cell
that would be coniferous, broadleaved, or unforested without any human
activity. Additionally we let $p_H$ denote the share of each grid cell
that is affected by ALCC. Since the ALCC data represents human land-use for food
production we assume that all human land-use can be seen as a replacement of
the corresponding proportion of natural land-cover with open land. The resulting
link between natural and actual land cover, $\mv{z}=(z_C,z_B,z_U)$, is
\begin{equation*}
  \begin{split}
    z_C &= p_C(1- p_H),\\
    z_B &= p_B(1- p_H),\\
    z_U &= p_U(1- p_H) + p_H,
  \end{split}
\end{equation*}
with the transformation being denoted $\mv{z}=h(\mv{p}_L, p_H)$ \citep[compare
to the covariate adjustments in][]{Pirzamanbein2014}.

The pollen based land-cover compositions $\mv{L} = (L_C,L_B,L_U)$ are now seen
as Dirichlet distributed observations of the actual land cover,
$\mv{z}$. Similarly the ALCC proportions $H$ are modelled as draws from beta
distributions with expectation $p_{H,k}$, where $p_{H,k}$ are 
perturbations of $p_{H}$ introduced to handle the (large) differences
between the two ALCC datasets (see Figure~\ref{figD:hyde_1400}). The resulting
model for the pollen and ALCC data given the underlying proportions is
\begin{equation}
  \begin{split}
    \mv{L}(\mv{s})|\alpha,\mv{z}(\mv{s}) &\sim \Dir{\alpha}{\mv{z}(\mv{s})},
    \\
    H_k(\mv{s})|\lambda, p_{H,k}(\mv{s}) &\sim
    \BetaDist{\lambda p_{H,k}(\mv{s})}{
      \lambda(1-p_{H,k}(\mv{s}))}.
  \end{split}
  \label{eqD:obs}
\end{equation}
Here $\mv{s}$ is the location of each grid cell and $\alpha$ and $\lambda$ are
concentration parameters controlling the uncertainty in the Dirichlet and beta
distributions.

We model the grid cell proportions $\mv{p}_L(\mv{s}) = (p_C(\mv{s}),\,
p_B(\mv{s}),\,p_U(\mv{s}))$ and $p_H(\mv{s})$ as a transformation
of an multivariate latent field $\mv{\eta}(\mv{s})$,
\begin{align*}
 \mv{p}_L(\mv{s}) &= f\bigl( \mv{\eta}_L(\mv{s}) \bigr), &  
 p_H(\mv{s}) &= g\bigl( \eta_H(\mv{s}) \bigr) 
\end{align*}
with $ f: \mathbb{R}^2 \rightarrow (0,1)^3$ and $g: \mathbb{R} \rightarrow
(0,1)$. For $f$ we use the inverse additive log-ratio transformation (applied
for each grid cell, $\mv{s}$),
\begin{equation}
  \begin{split}
    \mv{\eta}_L &= \left( \log\left(\dfrac{p_{C}}{p_{U}}\right),
      \log\left(\dfrac{p_{B}}{p_{U}}\right) \right) = 
    \left( \eta_{L_1}, \eta_{L_2} \right) 
    \\
    p_{\bullet} &= 
    \begin{cases} 
      \dfrac{\exp(\eta_{L_i})}{1+\sum_{i}\exp(\eta_{L_i})}
      &\text{for $p_C$ and $p_B$ with $i=1,2$,}
      \\\vspace{-0.3cm}&\\
      \dfrac{1}{1+\sum_{i}\exp(\eta_{L_i})} 
      &\text{for $p_U$.}
    \end{cases}
  \end{split}
  \label{eqD:alr}
\end{equation}
and for $g$ the inverse logit transformation
\begin{align}
  \eta_H &= \log\left( \dfrac{p_H}{1-p_H} \right)
  & &\text{and} &
  p_H &= \dfrac{\exp(\eta_H)}{1+\exp(\eta_H)}. 
  \label{eqD:logit}
\end{align}
The components of the latent field $\mv{\eta}(\mv{s})$ are collected into a
column vector and modelled using a mean part, $\mv{B}\mv{\beta}$,
and a component capturing spatial dependencies $\mv{X}$:
\begin{equation*}
  \begin{split}
    \mv{\eta} &= \begin{bmatrix}
      \mv{\eta}_{L_1} \\
      \mv{\eta}_{L_2} \\
      \mv{\eta}_H
      \end{bmatrix} = \mv{B}\mv{\beta}+ \mv{X}.
\end{split}
\end{equation*}
Here $\mv{B}$ is a matrix of covariates, $\mv{\beta}$ is a vector of regression
coefficients, and $\mv{X}$ is a multivariate spatial field.

For $\mv{\eta}_H$ covariates in $\mv{B}$ consist of an intercept and
elevation. For $\mv{\eta}_L$ two possible sets of covariates consisting of either
intercept and elevation; or intercept, elevation, and model based vegetation
estimates (from LPJ-GUESS) will be evaluated. For the LPJ-GUESS covariates the 
DVM based 3-compositions of natural potential vegetation were transformed to
$\mathbb{R}^2$ using \eqref{eqD:alr}, resulting in two covariates,
LPJ-GUESS$_{1,2}$.
The spatial field, $\mv{X}$, is modelled using a Gaussian Markov random field
\citep[GMRF,][]{rue2004gaussian} with a separable covariance structure,
\begin{equation*}
  \mv{X} \sim \Normal{\mv{0}}{\mv{\Sigma}\otimes \mv{Q}(\kappa)^{-1}}
\end{equation*}
where $\mv{\Sigma}$ is a $3\times 3$ covariance matrix, $\mv{Q}(\kappa)$ is
the precision matrix of a GMRF that approximates fields with Mat\'ern
covariance function \citep{lindgren2011explicit,lindgrenINLA}, and $\kappa$
governs the range of the spatial dependence.

To handle the differences between the KK10 and HYDE data, perturbed proportions
of human land-use $p_{H,k}(\mv{s})$ were introduced in the data model,
\eqref{eqD:obs}. These perturbations are created by adding random effects to the
$\mv{\eta}_H$-field;
$p_{H,k}(\mv{s})$ is computed from $\eta_{H,k}(\mv{s}) = \eta_H(\mv{s}) +
\epsilon_k$ using \eqref{eqD:logit} where $\epsilon_k \sim
\Normal{0}{\tau_\epsilon^{-1}}$. Note that $\epsilon_k$ are common
terms added to the entire field, an attempt to use different random effects for
each grid cell, i.e.\ $\epsilon_k(\mv{s})$, resulted in an unidentifiable model.

\begin{figure}[htp]
\centering
\begin{tikzpicture}[node distance=1cm,>=stealth',bend angle=45,auto]
  \tikzstyle{place}=[circle,thick,draw=blue!75,fill=blue!20,minimum size=5mm]
  \tikzstyle{transition}=[rectangle,thick,draw=black!75,fill=black!20,minimum size=5mm]
  \tikzstyle{trans}=[rectangle,rounded corners=1mm,inner sep=1mm,thick,draw=gray!75,fill=gray!20,minimum size=4mm]
  \tikzstyle{transitionw}=[rectangle,thick,draw=gray!75,fill=gray!20,minimum size=5mm]
  \tikzstyle{place_eta}=[ellipse,thick,draw=blue!75,fill=blue!20,minimum size=3mm]
  \begin{scope}
    
    \node [place] (kappa) {$\kappa$};

    \node [place] (rho) [right of=kappa]  {$\mv{\Sigma}$};

    \node [place] (phi) [left of=kappa] {$\varphi$};
    
    \node [place] (gamma) [left of=phi] {$\gamma$};
    
    \node [place] (alpha) [left of=gamma,xshift=-9mm] {$\alpha$};
    
    \node [place] (tau) [right of=rho,xshift=2.2mm]  {$\tau_\epsilon$};
	   
    \node [place] (lambda) [right of=tau,xshift=8mm] {$\lambda$};
    
 	\node [place_eta] (eta) [below of=kappa,xshift=-5mm] {$\mv{\eta}=\mv{B}\mv{\beta}+ \mv{X}$}
	edge [pre] (kappa)
    edge [pre] (rho)
    edge [pre] (phi)
    edge [pre] (gamma);
    
    \node [trans] (pH)  [below of=eta] {$p_H \equal g(\eta_H)$}
    edge [pre] (eta);
    
	\node [trans] (pL)  [left of=pH,xshift=-1.25cm] {$\mv{p}_L \equal f(\mv{\eta}_L)$}
    edge [pre] (eta);
    
    \node [trans] (pHk)  [right of=pH,xshift=1.8cm] {$p_{H,k} \equal g(\eta_H+\epsilon_k)$}
    edge [pre] (tau)
    edge [pre] (eta);
          	
    \node [trans] (z)  [below of=pL,xshift=1cm] {$\mv{z} \equal h(\mv{p}_L,p_H)$}
    edge [pre] (pH)
    edge [pre] (pL);
    
    \node [transition] (LCC)  [below of=z] {$\mv{L}$}
    edge [pre] (z);
          	
    \node [transition] (HLU)  [right of=LCC,xshift=3.05cm] {$\mv{H}_k$}
    edge [pre] (pHk);
    
    \draw[post] (alpha)|-(LCC);
    \draw[post] (lambda)|-(HLU);
 
  \end{scope}
\end{tikzpicture}
\caption{Directed acyclic graph describing the conditional dependencies in the
  hierarchical model.} 
\label{figD:hierarchical}
\end{figure}
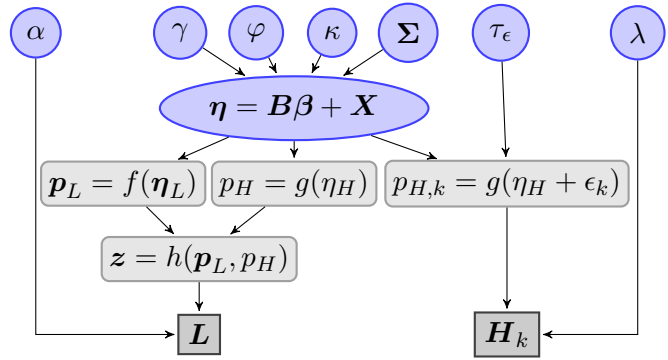

The full hierarchical model is illustrated in Figure~\ref{figD:hierarchical}.
The final part of the model is to specify suitable priors, following
\citep{pirzamanbein2018modelling} we use wide priors for $\alpha$ and $\lambda$;
conjugate priors for $\mv{\Sigma}$; for $\kappa$ we pick a
prior appropriate to the size of our spatial domain \citep{FuglsSLR2016_Interpretable}. For $\mv{\beta}$ we choose a grouped horseshoe shrinkage prior with global and local hyper-parameters $\varphi$ and $\gamma_i$. This shrinks insignificant coefficients towards zero aiding the variable selection. Hyper-parameters for $\gamma_i$ and $\varphi$ are given as the standard half-Cauchy distribution ($\proper{C}^+$) \citep{horseshoe}. Finally we pick a conjugate prior for $\tau_\epsilon$ since this, similar to
$\mv{\Sigma}$, allows for simple MCMC updates. The resulting priors are
\begin{equation*}
  \begin{aligned}
    \alpha &\sim \GammaDist{1.5}{0.1}, \qquad
    \lambda \sim \GammaDist{1.5}{0.1},  &\tau_\epsilon &\sim \GammaDist{1.5}{0.1},
    \\
    \kappa &\sim \GammaDist{1}{\frac{\log(100)}{\sqrt{8}}}, &
    \mv{\Sigma} &\sim \IW{\mathbb{I}}{10},
    \\
    \beta_{ki}|\gamma_i,\varphi &\sim \Normal{0}{\varphi^2\gamma_i^2},&i &= 1\cdots p\\
    \gamma_i &\sim \Cauchy{0}{1}, & \varphi &\sim \Cauchy{0}{1}.
  \end{aligned}
  \label{eqD:full}
\end{equation*}

\subsection{Estimation using MCMC} \label{secD:MCMC}
To estimate model parameters and reconstruct the latent field we use a
block-updated MCMC algorithm. In the first block the latent fields -- $\mv{\eta}$,
$\mv{\beta}$, and $\epsilon_k$ -- and the Dirichlet and beta concentration
parameters -- $\alpha$ and $\lambda$ -- are updated using a MALA proposal
\citep{MALA} and the conjugate posterior for $\tau_\epsilon$, $\gamma$ and $\varphi$. In the second block, we update the range parameter of the GMRF --
$\kappa$ -- using a random walk in log scale and the covariance matrix --
$\mv{\Sigma}$ -- using the conjugacy (conditioned on $\kappa$). Finally
$\tau_\epsilon$ is updated using the conjugate posterior. In each iteration the
MCMC alternates between these three blocks. To get the desired acceptance
rate we use an adaptive scheme \citep{AdaptiveMCMC} where the step size of the
MALA proposal and the random walk are adjusted to maintain $57\%$ and $40\%$
acceptance rate, respectively \citep{acc_rate_all}. This MCMC is an
extension of the implementation, for a simpler model, described by
\citet{pirzamanbein2018modelling}.

We ran $100\,000$ MCMC iterations with a burn-in sample size of $10\,000$ to
estimates the parameters of each model. The MCMC chain plots show convergence
and good mixing of the parameters.

\section{Results and discussion} \label{secD:results}
The reconstruction of human land-use, potential natural vegetation and land-cover compositions are shown in Figure~\ref{figD:full_1425CE} for the 1425 CE time period. The results for the other time periods are available in Appendix \ref{appD:results}. 
In general, the reconstructions capture the variability in the observed datasets. The human land-use reconstructions mostly capture the spatial patterns of KK10 while the amount of land-use is closer to HYDE. Moreover, the model with only elevation as covariates estimates slightly higher amounts of human land-use compared to the model also including LPJ-GUESS as covariates. 

\newcommand{\captiontext}[1]{The observation datasets (row 1) and the
  reconstructions using two different sets of covariates (row 2 and 3) for
  #1. From left to right: land-cover composition, natural land-cover, and human land-use.}

\begin{figure}[htp]
\noindent\makebox[\textwidth]{
  \includegraphics[width=\textwidth]{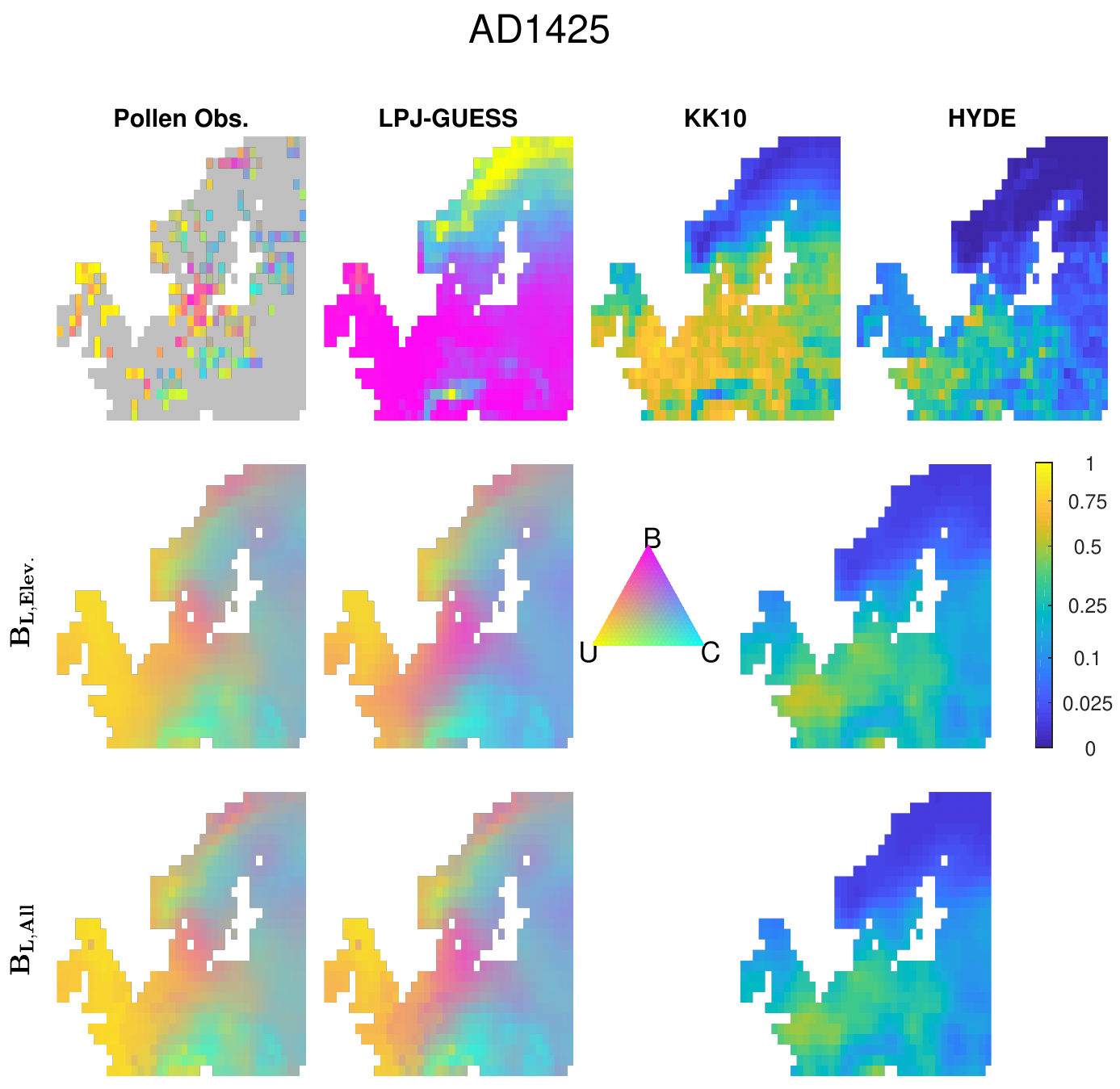}}
  \caption{\captiontext{1425 CE}}
  \label{figD:full_1425CE}
\end{figure}

\begin{figure}[htp]
\noindent\makebox[\textwidth]{
  \includegraphics[width=\textwidth]{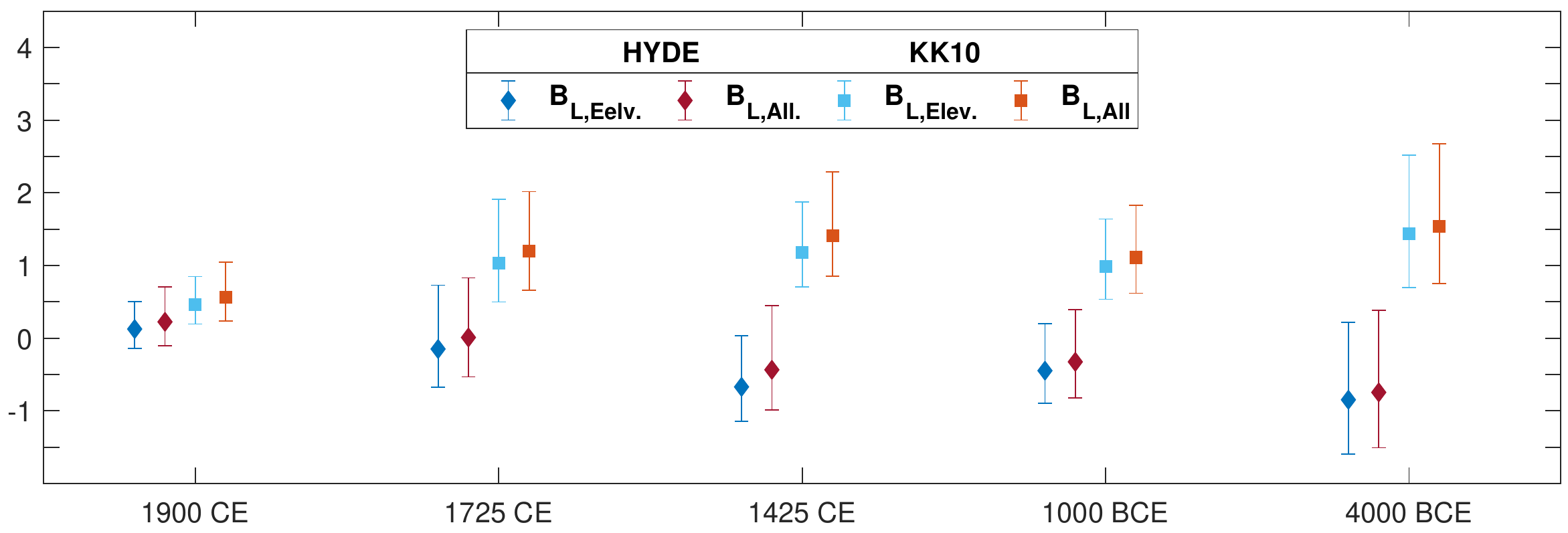}}
  \caption{Estimated $\epsilon_k$ for HYDE and KK10 and corresponding $95\%$ confidence intervals for all time periods. The blue color represents model includes only elevation and red color represents model include both elevation and LPJ-GUESS.}
  \label{figD:E_eps}
\end{figure}

The estimates of $\epsilon_k$ for HYDE and KK10 (Figure \ref{figD:E_eps}) also indicate that the human land-use reconstructions are, on average, closer to HYDE than KK10 for all the time periods. The difference between HYDE and
KK10, as captured by $\epsilon_k$, increases for older time periods (see Figure
\ref{figD:CI_HLU} in Appendix. \ref{appD:CI_HLU}). The estimates of $\epsilon_k$
are higher when the model includes both elevation and LPJ-GUESS as covariates
compared to the model only including elevation. This is in accordance with the
higher estimates of human land-use in the model containing only elevation.

\begin{figure}[htp]
\noindent\makebox[\textwidth]{
  \includegraphics[width=\textwidth]{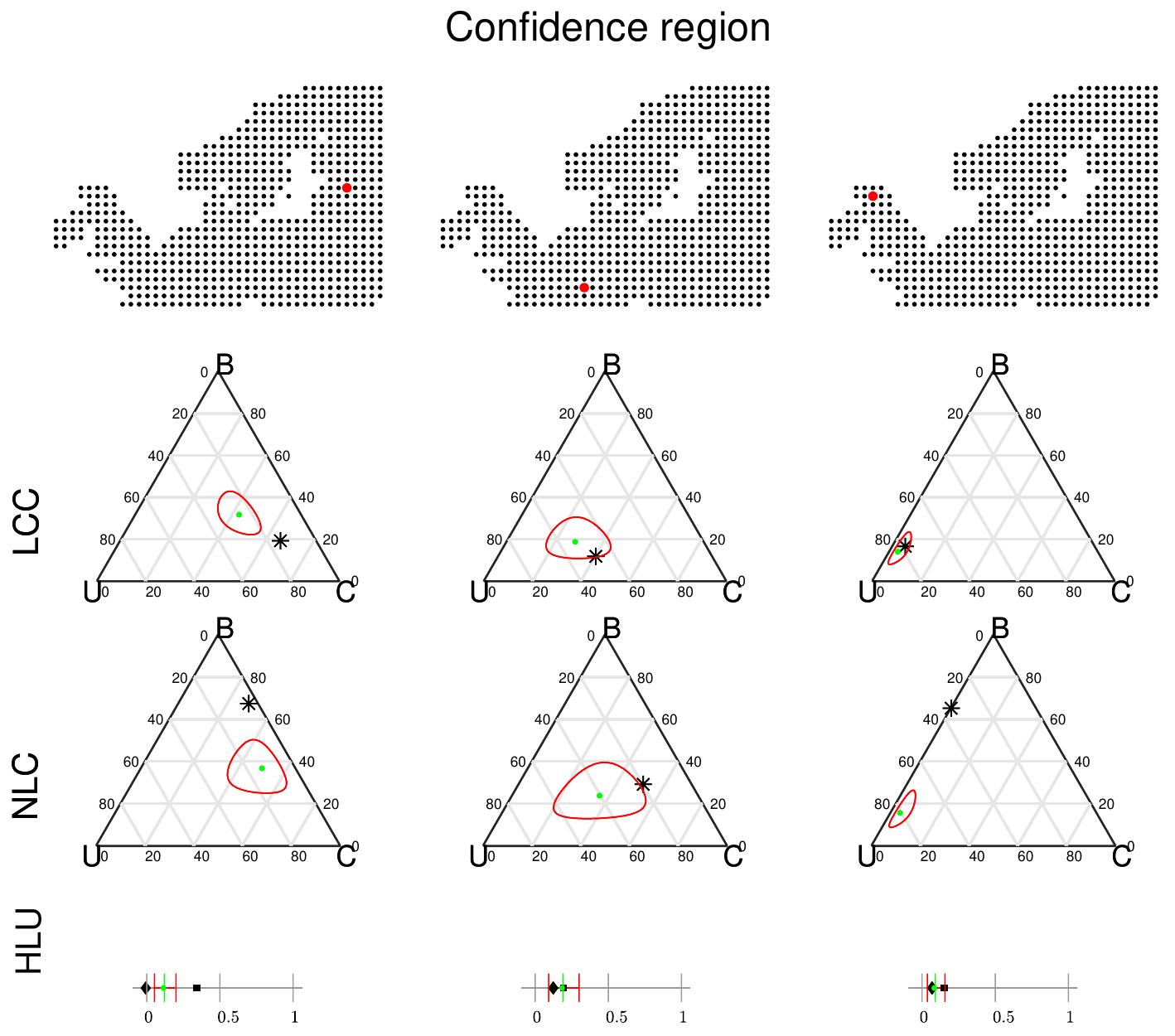}}
  \caption{The reconstruction and prediction regions for three locations for
    land-cover composition (LCC), natural land-cover (NLC) and human land-use
    (HLU) for 1425 CE. For LCC and NLC (rows 2 and 3) the observations, pollen
    based REVEALS reconstructions and LPJ-GUESS output respectively,  are marked
    with ($\ast$). For HLU (row 4) the two ALCC observations are given by HYDE
    (${\blacklozenge}$) and KK10 ({\tiny$\blacksquare$}). For all figures the
    green dots indicate estimated values and the red lines represent the
    corresponding confidence regions. All estimates and confidence regions are
    based on the  model without LPJ-GUESS$_{1,2}$ as covariates.}
  \label{figD:3_points_1425CE}
\end{figure}

The uncertainties in the human land-use reconstructions denote higher variation in the model with LPJ-GUESS as covariates than the model with only elevation (Figure \ref{figD:CI_HLU} in Appendix. \ref{appD:CI_HLU}).
The uncertainty in the compositional reconstructions, i.e. natural potential
land-cover and land-cover composition, are computed using transformed elliptical
confidence regions \citep{pirzamanbein2018modelling}. The results together with
confidence intervals for human land-use are illustrated in Figure~\ref{figD:3_points_1425CE} for three locations during the 1425 CE time period. The confidence regions are based on the model using only elevation, in order to allow a comparison between the NLC estimates and LPJ-GUESS. The selected point in the Baltic (column 1 in Figure \ref{figD:3_points_1425CE}) represents a location with contrasting values in the different data sources, i.e. about $70\%$ of coniferous forest in LCC, $70\%$ of broadleaved forest in LPJ-GUESS, and $40\%$ or $10\%$ of human land-use in KK10 and HYDE respectively. The differences among the data sources are balanced in the reconstruction of LCC, NLC and human land-use. In contrast, when the differences are smaller the confidence regions include the observations quiet well (columns 2 and 3 in Figure \ref{figD:3_points_1425CE}). The selected point in Scotland (column 3 in Figure \ref{figD:3_points_1425CE}) shows the improvement of the NLC reconstruction compared to the LPJ-GUESS estimate. The reconstruction suggests that the $80\%$ of unforested land consist of $10\%$ human land-use while LPJ-GUESS suggests $30\%$ unforested land and $70\%$ boardleaved forest.  

A leave out validation is used to evaluate the performance of the model. The
validation is performed by randomly removing $10\%$ of observed grid cells in
the LCC and ALCC data and reconstruct these values based on the remaining
observations. The resulting land-cover reconstructions are compared to LCC using
average compositional distance \citep[ACD;
see][]{aitchison2000logratio,pirzamanbein2018modelling}, and the human land-use
reconstructions are compared to both KK10 and HYDE using root mean
squared error (RMSE). Comparing the ACD and RMSE (Table \ref{tabD:LOO}), 
there is no general preference in for any of the two models with
different covariates. As has previously been noted the HLU estimates are, in
general, closer to HYDE than to KK10.

\begin{table}[htp]
{\renewcommand{\arraystretch}{1.5}%
\centerline{
\begin{tabular}{c|c|c||c|c|c|c|}
\multicolumn{1}{l}{}&\multicolumn{2}{c}{\textbf{ACD}}&\multicolumn{4}{c}{\textbf{RMSE}}\\
\cline{2-7}
\multicolumn{1}{l|}{}&\multicolumn{2}{c||}{REV}&\multicolumn{2}{c|}{KK10}&\multicolumn{2}{c|}{HYDE}\\
\multicolumn{1}{l|}{}&\multicolumn{1}{c|}{$\mv{B}_\text{All}$}&\multicolumn{1}{c||}{$\mv{B}_\text{Elev.}$}&\multicolumn{1}{c|}{$\mv{B}_\text{All}$}&\multicolumn{1}{c|}{$\mv{B}_\text{Elev.}$}&\multicolumn{1}{c|}{$\mv{B}_\text{All}$}&\multicolumn{1}{c|}{$\mv{B}_\text{Elev.}$}\\
\cline{2-7}
\textbf{1900 CE}	&	1.11	&	\textbf{0.97}	&	0.16	&	0.13	&	0.14	&	\textbf{0.12}\\
\cline{2-7}
\textbf{1725 CE}	&	1.11	&	\textbf{1.01}	&	0.26	&	0.22	&	0.14	&	\textbf{0.12}\\
\cline{2-7}
\textbf{1425 CE}	&	1.25	&	\textbf{1.15}	&	0.25	&	0.23	&	\textbf{0.09}	&	0.10\\
\cline{2-7}
\textbf{1000 BCE}   &	1.17	&	\textbf{1.14}	&	0.14	&	0.13	&	\textbf{0.05}	&	0.06\\
\cline{2-7}
\textbf{4000 BCE}   &	\textbf{1.14}	&	1.27	&	0.14	&	0.12	&	\textbf{0.02}	& \textbf{0.02}\\
\cline{2-7}
\end{tabular}}}
\caption{Leave out validation results for models with two different sets of covariates, $\mv{B}_\text{All}$,$\mv{B}_\text{Elev.}$ and all time periods. The reconstructions of land-cover compositions (LCC) are compared using average compositional distances (ACD). The human land-use (HLU) reconstructions  are compared using root mean square error (RMSE). The bold number indicates the lowest value in the row for LCC and HLU.}
\label{tabD:LOO}
\end{table}

\section{Conclusion}
\label{secD:conclusion}

In this paper, we developed a Bayesian hierarchical model to reconstruct the past human land-use for five time periods centred around 1900 CE, 1725 CE, 1425 CE, 1000 BCE and 4000 BCE. The reconstructions are based on combination of pollen based land-cover compositions \citep{Trondman2015} and population based anthropogenic land-cover changes (ALCC) estimates. 

Due to discrepancies between the past ALCC estimates, the model uses two
different datasets of human land-use: \begin{inparaenum}[1)] anthropogenic land
  cover changes scenario of \citet[][KK10]{kaplan2009prehistoric} and historic data
  base of global environment \citep[HYDE;][]{klein2011hyde}\end{inparaenum}. 
The past human land-use reconstruction capture the spatial patterns of KK10
while being closer in value to the proportions of HYDE. This suggests that
pollen based LCC can be used to adjust the existing population based human land
use to match observed past vegetation patterns and recover past human land-use
from pollen based LCC.

We note that the model would allow the inclusion of additional anthropogenic
land-cover changes scenarios and it would be interesting to also include
archaeological data. However, our initial attempts to use archaeological data
have so far, as described below, been unsuccessful.


\subsection{Including archaeological data in the model} \label{secD:archaeological}
We initially considered using archaeological data, instead of the ALCC
scenarios, as a measure of human land-use. Given an archaeological dataset
containing the 
locations of relevant archaeological finds during each of the five time periods
we would replace the $\beta$-observations of the ALCC scenarios with a point
process \citep{SimpsILSR2016_103} over the archaeological finds. The base idea
being that more finds, in a given region, would correspond to a
higher human activity and thus a higher proportion of ALCC.

One possible model would be an exponential link-function between the latent
field, $\mv{\eta}_H$, and the intensity, $\mv{\lambda}$, of the point process
for the archaeological finds, e.g. 
\begin{align*}
  \mv{\lambda} &= \exp\left( \mv{\eta}_H  \right)
  \\
  \log \pP( \mv{A} | \lambda) &= \abs{\Omega} - \int_\Omega \lambda(s)\md s
  + \sum_{i=1}^n \log\lambda(s_i)
\end{align*}
where $\mv{A}=\{s_i\}$ are the locations of the archaeological finds.
Since the point process provides the relative frequency of events, the
latent field, $\mv{\eta}_H$, might only be determined up to an additive constant.
To make the model identifiable the ALCC scenarios could still be needed, either
as observations or as covariates. While it would be very interesting to investigate this model we have been unable to find a suitable archaeological dataset.

For us, a large detrimental factor to the use of archaeological data has been our inability to find archaeological databases covering the entire study area.
One option considered was to restrict the modelling to Sweden using the
\textit{Forns\"{o}k}-database\footnote{http://www.raa.se/in-english/about-fornsok/} maintained by the Swedish National Heritage Board. This database contains information regarding roughly 1.7 million finds, but is incomplete with data contributions largely depending on the local municipalities
(\textit{kommuner}).

\begin{figure}[htp]
  \includegraphics[width=0.8\textwidth]{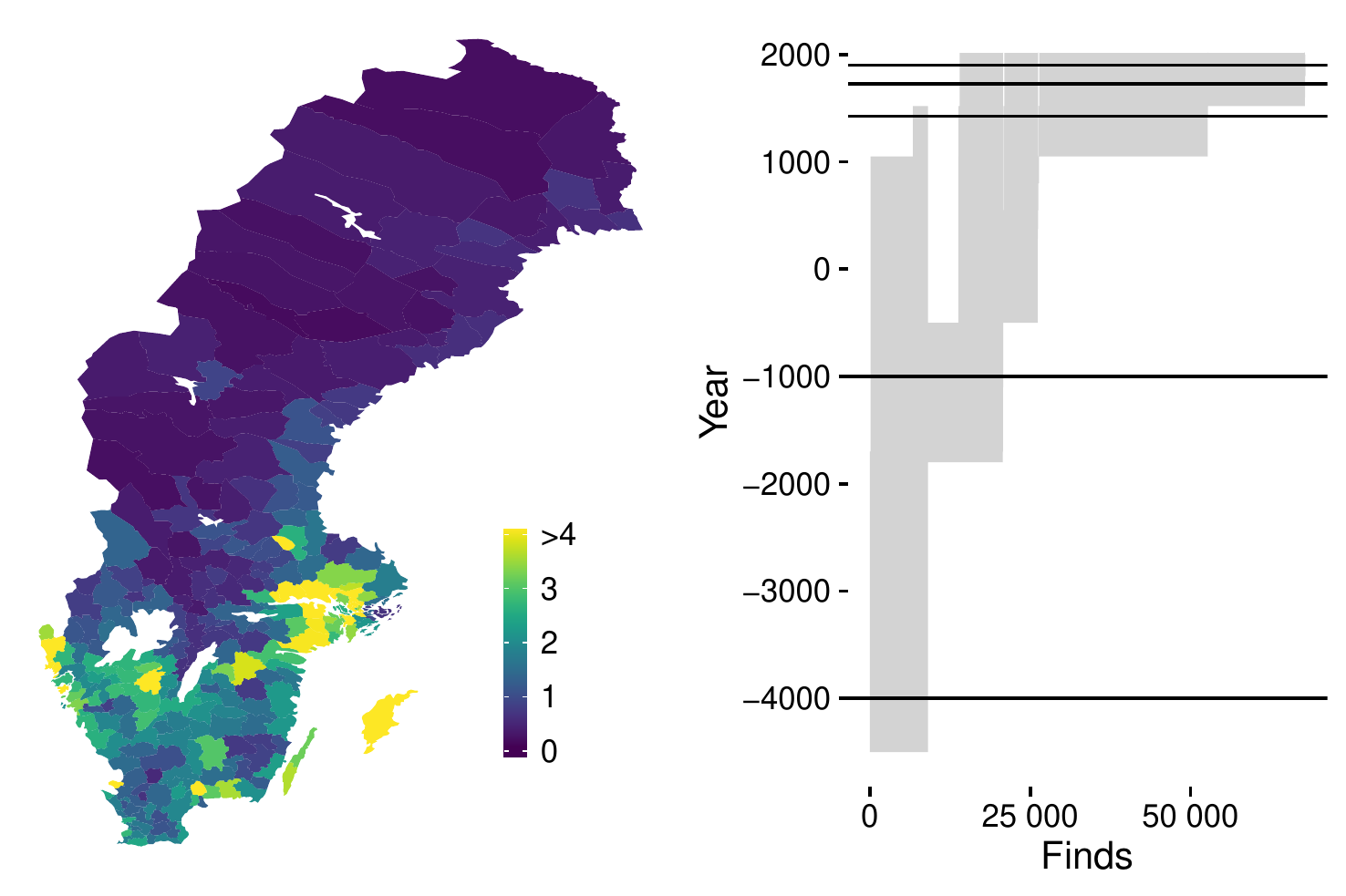}
  \caption{Overview of the Swedish archaeological data. The left pane shows the
    total number of finds per square kilometer for each of the 290
    municipalities (\textit{kommuner}) of Sweden. In the right pane the grey
    area indicates the dating range given for each archaeological find. The five
    time periods for which we have pollen data are indicated by the horizontal
    black lines.}
  \label{figD:arch_data}
\end{figure}

An initial search of the database resulted in $68\,000$ dated finds marked as relating to agricultural and/or settlement activities. And an additional $54\,000$ finds in these categories without any dating information.  The spatial information regarding finds is good ($\pm 250$ m, i.e.\ much smaller than the spatial resolution of the pollen based LCCs). However, the dating information ranges from very good (based on $C_{14}$ or dendrochronology) to rather inexact. With most of the finds being dated based on typology, i.e.\ as belonging to one (or several) of 5 time periods. The wide ranges of possible dates and the uncertainty regarding selection bias due to differing priorities among the contributing municipalities makes the data unsuitable for our purposes (see Fig.~\ref{figD:arch_data}).

\section*{Acknowledgement}
The research presented in this paper is a contribution to the
two Swedish strategic research areas Biodiversity and Ecosystems
in a Changing Climate (BECC), and ModElling the Regional
and Global Earth system (MERGE).

We thank M.-J. Gaillard and A. Poska for providing the pollen based land-cover data complied by LAND Cover-CLIMate interactions in NW Europe during the Holocene (LANDCLIM) project, natural vegetation cover from LPJ-GUESS, and anthropogenic land-cover changes of KK10 data bases.

\clearpage

\appendix
\renewcommand{\thesection}{\Alph{section}}

\section{Computation for MALA proposal}
For MALA proposal, the computation of the log density, first derivatives  and
expected Fisher information of the Beta distribution are required. The Fisher information
is the negative expectation with respect to observations of the second and partial derivatives of the log density with respect to parameters and latent field.

\subsection{Beta distribution computations}
The Beta density is 
\begin{align*}
  \pP(\mv{y}|\lambda,\mv{p}) &= \frac{\Gamma(\lambda)}{
    \Gamma(\lambda\mv{p})\Gamma(\lambda(1-\mv{p}))} 
  \mv{y}^{\lambda\mv{p}-1}(1-\mv{y})^{\lambda(1-\mv{p})-1} 
  &
  \lambda&>0,\ \mv{p}\in(0,1),
\end{align*}
therefore the log density becomes
\begin{equation*}
\begin{split}
l = \log \pP(\mv{y}|\lambda,\mv{p}) =& \log\Gamma(\lambda) - \log\Gamma(\lambda\mv{p}) -\log\Gamma(\lambda(1-\mv{p}))\\ 
&+ (\lambda\mv{p}-1)\log \mv{y} + (\lambda(1-\mv{p})-1)\log (1-\mv{y}). 
\end{split}
\end{equation*}
The first derivatives with respect to the parameters, $\lambda$ and $\mv{p}$ are
\begin{equation*}
\begin{split}
\dfrac{\partial l}{\partial\lambda} &= \psi(\lambda)- \mv{p}\psi(\lambda\mv{p})-(1-\mv{p})\psi(\lambda(1-\mv{p}))+\mv{p}\log \mv{y}+ (1-\mv{p})\log(1- \mv{y}),\\
\dfrac{\partial l}{\partial\mv{p}} &= -\lambda\psi(\lambda\mv{p})+\lambda\psi(\lambda(1-\mv{p}))+\lambda\log\mv{y}-\lambda\log(1-\mv{y}).
\end{split}
\end{equation*}
The second and partial derivatives are
\begin{equation*}
\begin{split}
\dfrac{\partial^2 l}{\partial\lambda^2} = &\psi'(\lambda)- \mv{p}^2\psi'(\lambda\mv{p})-(1-\mv{p})^2\psi'(\lambda(1-\mv{p})),\\
\dfrac{\partial^2 l}{\partial\mv{p}^2} = &-\lambda^2\psi'(\lambda\mv{p})-\lambda^2\psi'(\lambda(1-\mv{p})),\\
\dfrac{\partial^2 l}{\partial\mv{p}\partial\lambda} = &- \psi(\lambda\mv{p})-\lambda\mv{p}\psi'(\lambda\mv{p})+\psi(\lambda(1-\mv{p}))+\lambda(1-\mv{p})\psi'(\lambda(1-\mv{p}))\\
&+\log \mv{y}-\log(1- \mv{y}).
\end{split}
\end{equation*}
The symmetric Fisher information is
\begin{equation*}
\mathcal{I} = \bmat{\mathcal{I}_{\lambda,\lambda}&\mathcal{I}_{\lambda,\mv{p}}\\&\mathcal{I}_{\mv{p},\mv{p}}}= -\pE_\mv{y} \bmat{\dfrac{\partial^2 l}{\partial\lambda^2}&\dfrac{\partial^2 l}{\partial\mv{p}\partial\lambda}\\&\\
&\dfrac{\partial^2 l}{\partial\mv{p}^2}}
\end{equation*}
with elements
\begin{equation*}
\begin{split}
\mathcal{I}_{\lambda,\lambda} = &-\psi'(\lambda)+ \mv{p}^2\psi'(\lambda\mv{p})+(1-\mv{p})^2\psi'(\lambda(1-\mv{p}))\\
\mathcal{I}_{\mv{p},\mv{p}} = &\lambda^2\psi'(\lambda\mv{p})+\lambda^2\psi'(\lambda(1-\mv{p}))\\
\mathcal{I}_{\lambda,\mv{p}} = &-\psi(\lambda\mv{p})-\lambda\mv{p}\psi'(\lambda\mv{p})+\psi(\lambda(1-\mv{p}))+\lambda(1-\mv{p})\psi'(\lambda(1-\mv{p}))\\
 &+\log\mv{y} -\log(1-\mv{y}).
\end{split}
\end{equation*}
Since $\pE(\log \mv{y}) = \psi(\lambda\mv{p})-\psi(\lambda)$, $\mathcal{I}_{\lambda,\mv{p}}$ simplifies to
\begin{equation*}
\begin{split}
\mathcal{I}_{\lambda,\mv{p}} = \lambda\mv{p}\psi'(\lambda\mv{p}) - \lambda(1-\mv{p})\psi'(\lambda(1-\mv{p})).
\end{split}
\end{equation*}


\clearpage
\section{Maps of reconstructed land-cover and human land-use}
\label{appD:results}

\begin{figure}[htp]
\noindent\makebox[\textwidth]{
  \includegraphics[width=\textwidth]{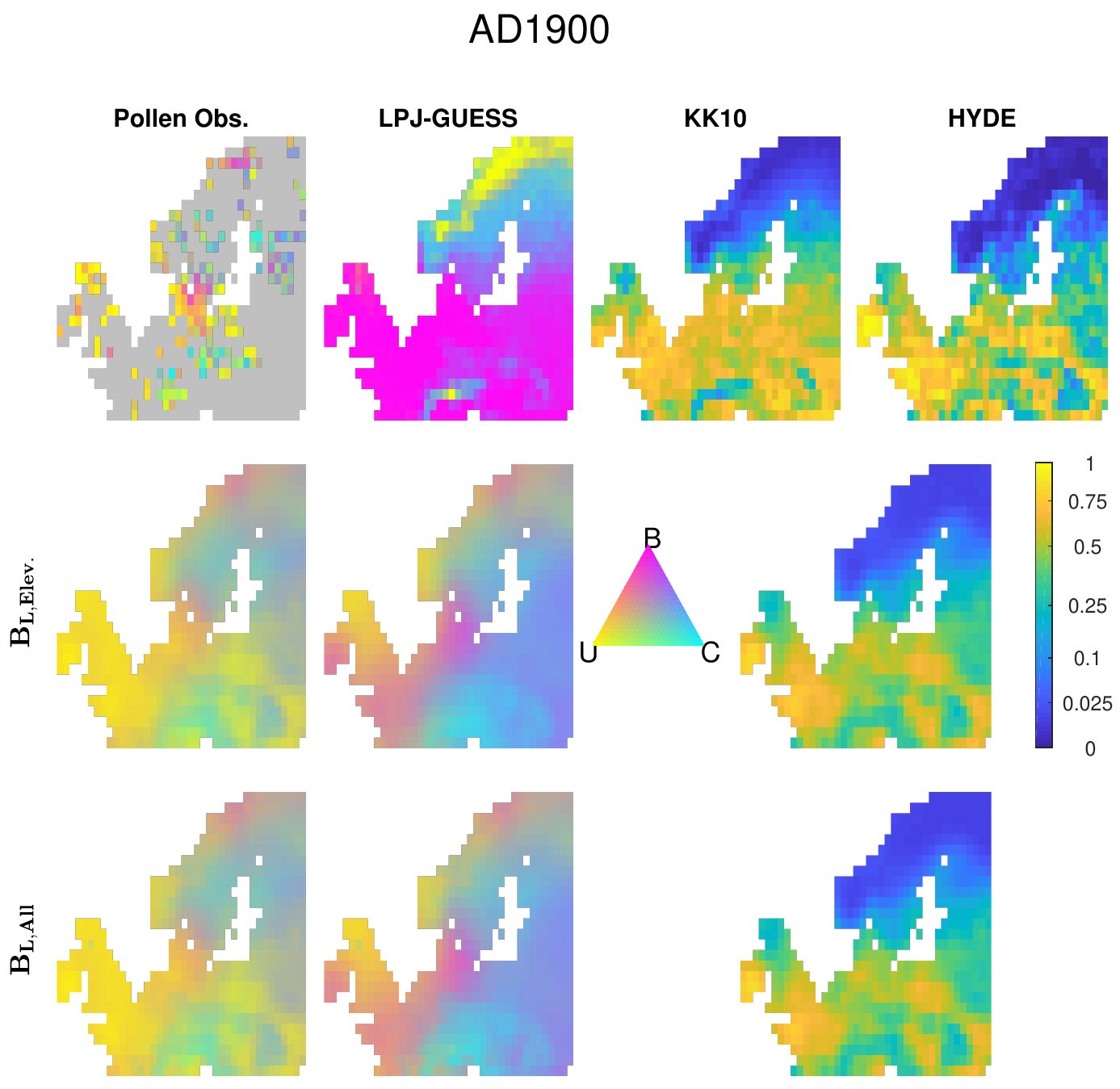}}
  \caption{\captiontext{1900 CE}}
  \label{figD:full_1900CE}
\end{figure}

\begin{figure}[htp]
\noindent\makebox[\textwidth]{
  \includegraphics[width=\textwidth]{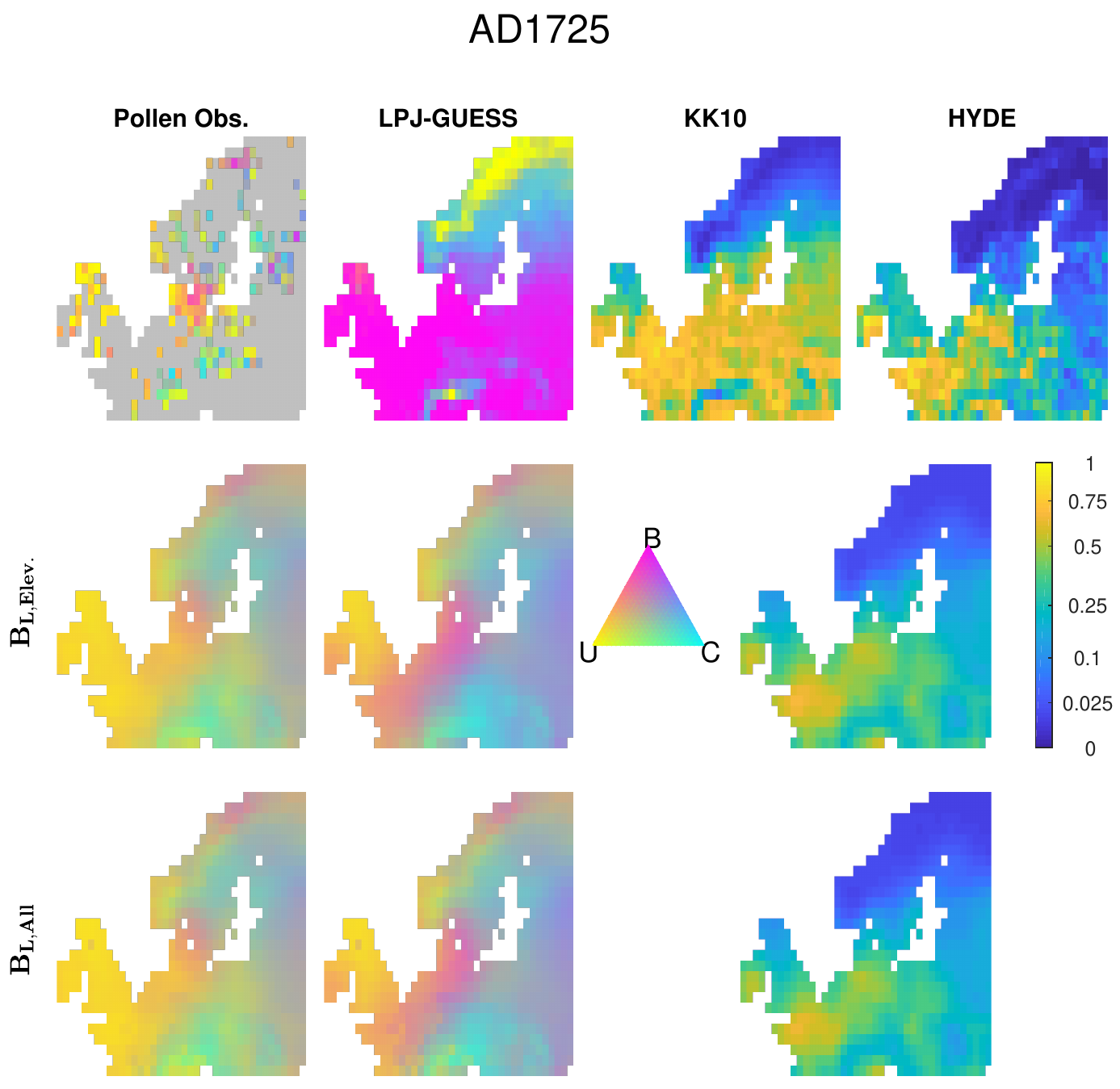}}
  \caption{\captiontext{1725 CE}}
  \label{figD:full_1725CE}
\end{figure}

\begin{figure}[htp]
\noindent\makebox[\textwidth]{
  \includegraphics[width=\textwidth]{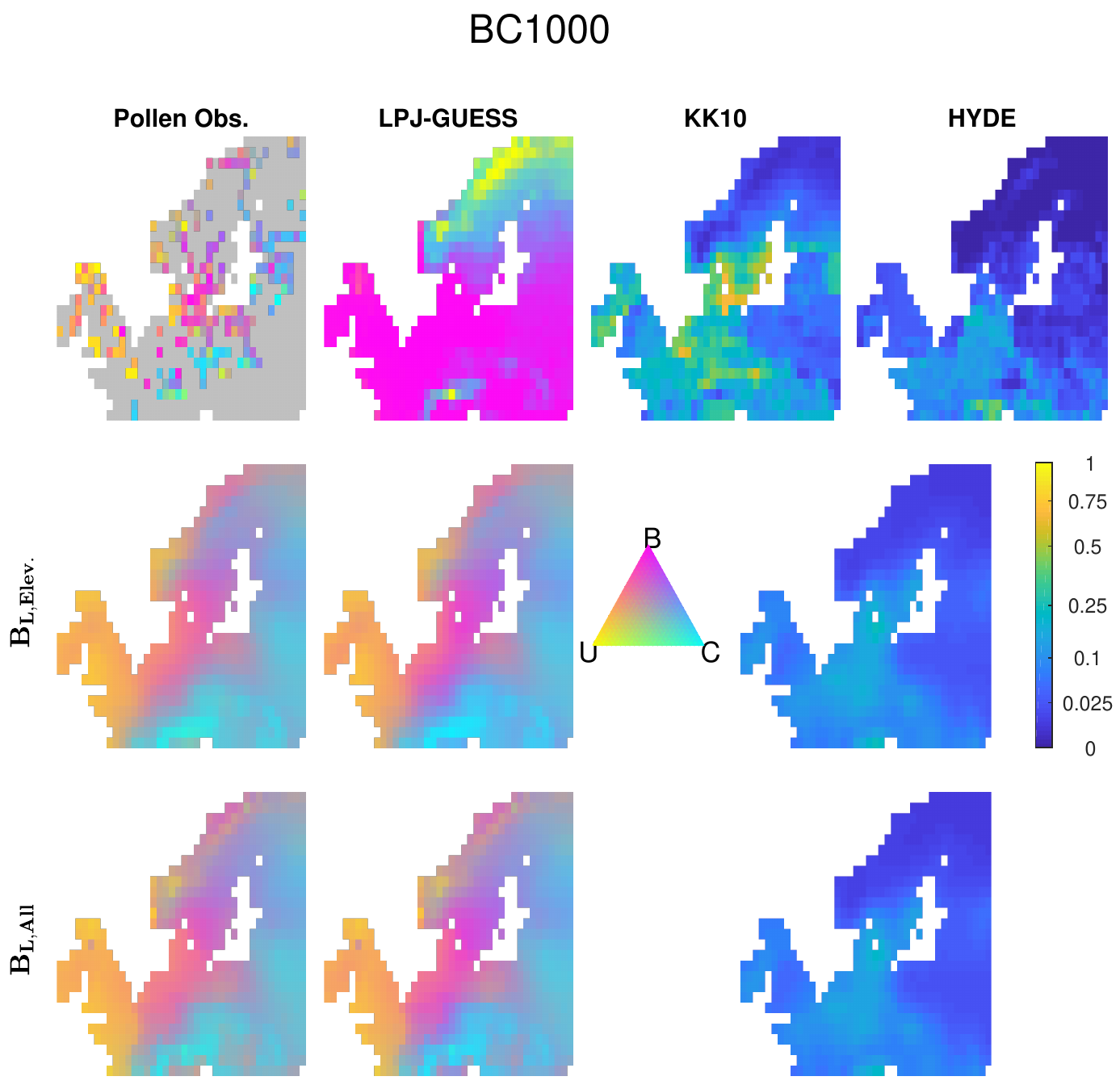}}
  \caption{\captiontext{1000 BCE}}
  \label{figD:full_1000BCE}
\end{figure}

\begin{figure}[htp]
\noindent\makebox[\textwidth]{
  \includegraphics[width=\textwidth]{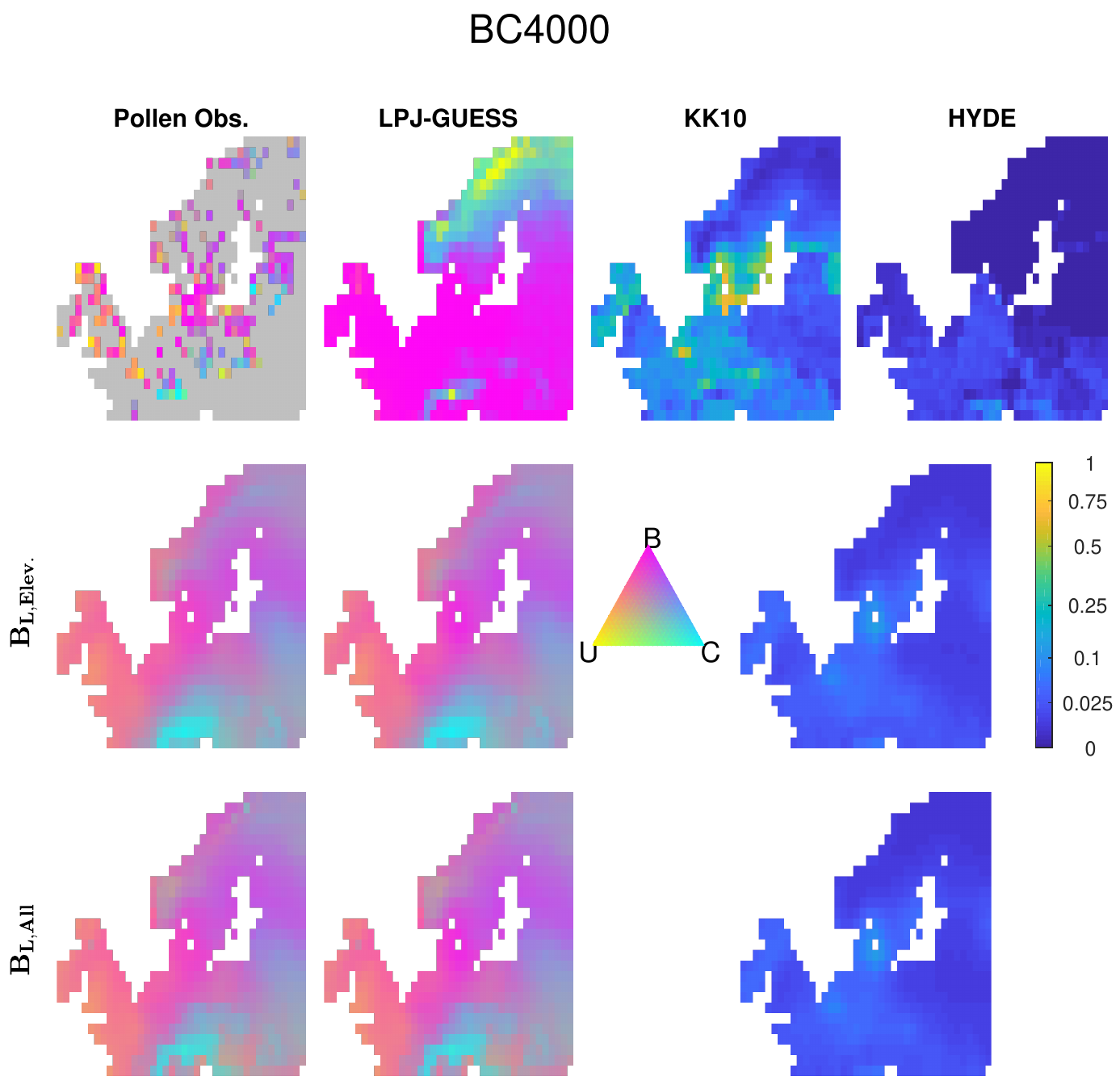}}
  \caption{\captiontext{4000 BCE}}
  \label{figD:full_4000BCE}
\end{figure}
\clearpage

\section{Uncertainties in land-use reconstruction}
\label{appD:CI_HLU}
The desription of the figure in this appendix is as follows,
\\
$95\%$ confidence interval for human land-use reconstructions for all time periods. From left to right: HYDE observations, KK10 observations, lower bound and upper bound for reconstruction of the model with only elevation as covariates, $\mv{B}_\text{Elev.}$, and lower bound and upper bound for reconstructions of the model with both elevation and LPJ-GUESS as covariates, $\mv{B}_\text{All}$.

\begin{sidewaysfigure}
\noindent\makebox[\textwidth]{
  \includegraphics[width=\textwidth]{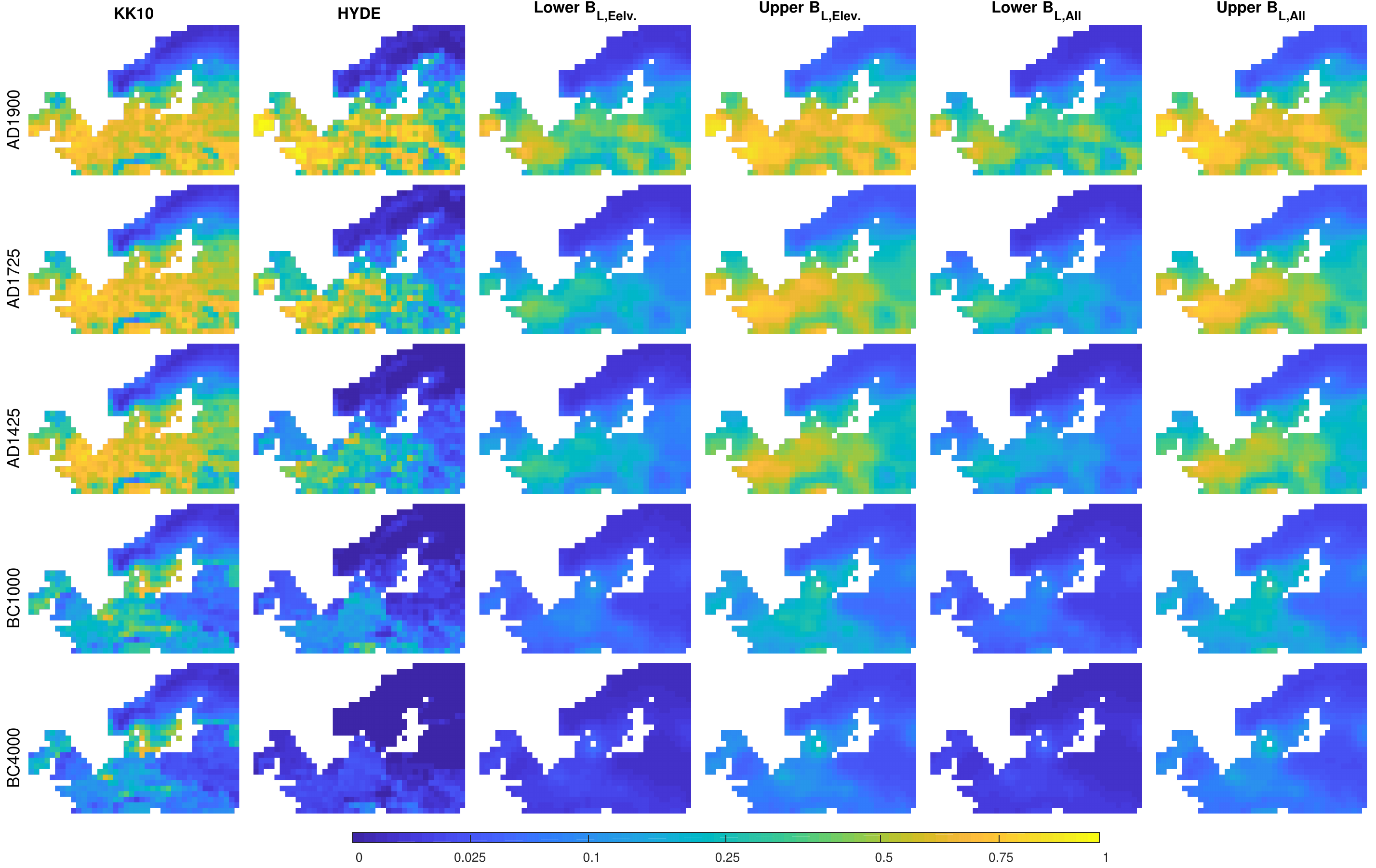}}
  \caption{HYDE and KK10 observations and confidence bound for human land-use reconstructions, see page \pageref{appD:CI_HLU}.}
  \label{figD:CI_HLU}
\end{sidewaysfigure}

\bibliographystyle{abbrvnat}
\bibliography{Journals_abrv,third_bib}

\end{document}